\documentclass[amssymb,aps,nofootinbib,letterpaper,10pt,twocolumn,nobalancelastpage]{revtex4-2}

\usepackage{graphicx}% Include figure files
\usepackage{dcolumn}% Align table columns on decimal point
\usepackage{bm}% bold math
\usepackage{amsmath}
\usepackage{amsfonts,amssymb}
\usepackage{algorithmic}
\usepackage{graphicx}
\usepackage{epsfig}
\usepackage{booktabs}
\usepackage{dsfont}
\usepackage[position=top]{subfig}
\usepackage{tabularx}
\usepackage{tikz}
\usepackage{dot2texi}
\usetikzlibrary{shapes,arrows}
\usetikzlibrary{backgrounds,calc,decorations.pathreplacing}
\usepackage[pdf]{graphviz}

\usepackage{mathtools}
\usepackage{multirow}
\usepackage{color}
\usepackage{xcolor}
\usepackage{footnote}
\newtheorem{example}{Example}

\usepackage[linesnumbered,ruled,vlined]{algorithm2e}
\usepackage{caption}
\usepackage{microtype}
\usepackage[utf8]{inputenc}
\usepackage{tikz}
\usetikzlibrary{snakes,arrows,shapes}
\usepackage{amsmath}
\usepackage{tabularx}
\usepackage{multirow}
\usepackage{pgfplots}
\usetikzlibrary{positioning,shapes,arrows,shadows,patterns}
\usepackage{soul}
\usepackage{xfrac}  
\usepackage{flushend}

\begin{document}

%\preprint{APS/123-QED}

\title{Efficient Deterministic Preparation of Quantum States \\Using Decision Diagrams}% Force line breaks with \\

\author{Fereshte Mozafari$^1$}
 \email{fereshte.mozafari@epfl.ch}
\author{Giovanni De Micheli$^1$}
\homepage{\url{https://si2.epfl.ch/~demichel/}}
\author{Yuxiang Yang$^{2,3}$}
 \email{yuxiang@cs.hku.hk}

\affiliation{%
\medskip
\footnotesize
$^1$\textit{\mbox{Integrated Systems Laboratory, EPFL, Lausanne, Switzerland}}\\%
$^2$\textit{\mbox{QICI Quantum Information and Computation Initiative, Department of Computer Science,} \mbox{The University of Hong Kong, Pokfulam Road, Hong Kong, China} }\\
$^3$\textit{\mbox{Institute for Theoretical Physics, ETH Z\"{u}rich, Z\"{u}rich, Switzerland} }\\% Department of Computer Science,
}%

\begin{abstract}
Loading classical data into quantum registers is one of the most important primitives of quantum computing. While the complexity of preparing a generic quantum state is exponential in the number of qubits, in many practical tasks the state to prepare has a certain structure that allows for faster preparation. In this paper, we consider quantum states that can be efficiently represented by (reduced) decision diagrams, a versatile data structure for the representation and analysis of Boolean functions. We design an algorithm that utilises the structure of decision diagrams to prepare their associated quantum states. Our algorithm has a circuit complexity that is linear in the number of paths in the decision diagram. 
Numerical experiments show that our algorithm reduces the circuit complexity by up to 31.85\% compared to the state-of-the-art algorithm, when preparing generic $n$-qubit states with $n^3$ non-zero amplitudes.
Additionally, for states with sparse decision diagrams, including the initial state of the quantum Byzantine agreement protocol, our algorithm reduces the number of CNOTs by 86.61\% $\sim$ 99.9\%.
\end{abstract}

\maketitle

\section{\label{sec:intro}Introduction}
Quantum computers are expected to provide advantages in several fields such as optimization~\cite{bravyi2018quantum},  chemistry~\cite{hempel2018quantum}, machine learning~\cite{biamonte2017quantum}, and materials science~\cite{aspuru2005simulated}. However, the quantum speedups can be sabotaged if the cost of loading data and initialization is too high for the quantum computer~\cite{biamonte2017quantum}. Therefore, minimising the cost of \emph{quantum state preparation} (QSP), the process of preparing quantum states from their classical descriptions, is a crucial step of quantum computation~\cite{plesch2011quantum,mottonen2004transformation,ItenCK2016}.

QSP algorithms for preparing \emph{general $n$-qubit quantum states} have cost that grows exponentially fast in $n$ ~\cite{Mottonen04, Shende06, Kaye04, Niemann16, iten2016quantum}.
Here the cost is quantified by the number of required CNOT gates, as
any quantum circuit can be decomposed into CNOT gates and single-qubit gates and the number of single-qubit gates is upper bounded by twice the number of CNOTs~\cite{shende2004minimal}. In this work, we focus on algorithms that prepare quantum states in a deterministic manner with no or fixed ancillary qubit overhead, instead of approximate algorithms~\cite{soklakov2006efficient, sanders2019black, zoufal2019quantum} or algorithms with $n$-dependent ancilla size~\cite{araujo2020divide, babbush2018encoding, zhang2022quantum}.

In contrast to general quantum states, in most quantum computational tasks, the states to prepare are from subfamilies of $n$-qubit states, such as uniform quantum states~\cite{mozafari2021dd, mozafari2020automatic}, Dicke states~\cite{bartschi2019deterministic}, and cyclic quantum states~\cite{mozafari2022cyclic}.
In these examples, all state subfamilies have classical descriptions with symmetric structures, which hints at the possibility of utilizing structured classical descriptions of quantum states to achieve efficient QSP. Here we exploit this possibility and propose a novel QSP algorithm for quantum states represented by reduced ordered decision diagrams. 
Decision diagrams are directed acyclic graphs over a set of Boolean variables and a non-empty terminal set with exactly one \emph{root} node~\cite{bahar1997algebric}. Decision diagrams avoid redundancies and lead to a more compact representation of logic functions. 

In this work, we consider the preparation of $n$-qubit quantum states $|\varphi\rangle = \sum_{s\in S} {\alpha_{s}} |s\rangle$, i.e., finding a unitary circuit $U$ that consists of elementary quantum gates such that $U|0\rangle^{\otimes n}= |\varphi\rangle$.
Here the \emph{index set} $S\subset\{0,1\}^{n}$ contains every binary string $s$ such that the amplitude $\alpha_s$ of $|s\rangle$ is non-zero, and $\sum_{s\in S} |\alpha_{s}|^2 = 1$. %We denote by $m:=|S|$ the cardinality of $S$.
Without loss of generality, we assume that basis states in $S$ are sorted in descending order. For two arbitrary $n$-bit strings $s$ and $s'$, there is a natural order $s\succ s'$ if $s$ is no smaller than $s'$ when both are regarded as binary numbers. In this way, we can order the elements of $S$ as $s_1\succ s_2\succ\cdots\succ s_m$ and express the state to prepare as
\begin{equation}\label{state-to-prepare}
    |\varphi\rangle = \sum_{i=1}^m {\alpha_{s_i}} |s_i\rangle.
\end{equation}

We use decision diagrams to represent the state in Eq.~(\ref{state-to-prepare}), where each basis state $|s_i\rangle$ and each amplitude $\alpha_{s_i}$ are represented by a path and a terminal node, respectively.
We propose an efficient algorithm that prepares an arbitrary quantum state given its associated decision diagrams. 
%To prepare basis states inorder, we traverse the decision diagram in pre-order traversal. 
The cost of our algorithm is $O(kn)$, where $k$ is the number of paths in the decision diagram. Since $k$ is always upper bounded by (and can be much smaller than) $m$, the number of non-zero amplitudes of the state in the computational basis, our algorithm efficiently prepares any \emph{sparse state} with $m\ll 2^n$. 
Sparse quantum states have many applications for example in quantum linear system solvers~\cite{harrow2009quantum}, quantum Byzantine agreement algorithm~\cite{ben2005fast} for large $n$, and quantum machine learning~\cite{biamonte2017quantum}. 
%There are many quantum algorithms that require special initial quantum states to encode a problem instance and depending on the problem instance, these states can also be sparse. 
Besides, many problems in classical computing are sparse such as sparse (hyper) graph problems~\cite{streinu2009sparse}. To solve them using a quantum computer, we need to prepare their associated sparse quantum states. In addition, our algorithm can also efficiently prepare states with sparse decision diagrams ($k \ll 2^n$), even if the states themselves are not sparse ($m=\Omega(2^n)$). 

Several algorithms have been proposed for sparse quantum state preparation~\cite{de2021double, malvetti2021quantum,sparseQSP} with $O(mn)$ cost. In all of them, the idea is based on preparing basis states one-by-one by applying several CNOTs and one multiple-controlled single-target gate. 
Tiago et al.~\cite{de2021double} use one ancilla qubit to avoid disturbing prepared basis states while working on the others.  Compared to ~\cite{malvetti2021quantum}, their results show that their algorithm performs well when the number of 1 bits in binary bit string representation of each basis state is almost 20\%, which is a limitation. 
Emanuel et al.~\cite{malvetti2021quantum} propose an algorithm to prepare sparse isometries which include sparse states as well. Niels et al.~\cite{sparseQSP} propose an algorithm that works in the opposite direction, i.e., they try to apply some gates to obtain $|0\rangle^{\otimes n}$ state from the desired sparse state. They repeat the same procedure in $m$ iterations. In every iteration, they select two basis states and merge them into one by applying several CNOTs and one multiple-controlled single-target gate. Comparing methods in~\cite{malvetti2021quantum} and~\cite{sparseQSP}, they both perform well with small $m$, and their circuit costs are almost the same. However, the idea in~\cite{sparseQSP} is simpler and its classical runtime, which is $O(n m \log_2(m))$, is less than that of the algorithm in ~\cite{malvetti2021quantum}, which is $O(\binom{n}{\log_2(m)} + n m^2 )$. Hence, we regard~\cite{sparseQSP} as the state of the art and compare our results to it.

Numerical experiments show that our algorithm outperforms the state of the art~\cite{sparseQSP}. Depending on the sparsity $m$, our algorithm achieves an up to 31.85\% reduction of the CNOT cost. The algorithm works very well for the states with sparse decision diagram representations, and uses up to $99.97\%$ fewer CNOTs. In addition, our algorithm requires only one ancilla qubit, in stark contrast to many existing works \cite{araujo2020divide, babbush2018encoding, zhang2022quantum} with ancilla qubit that grow with $n$.

\section{\label{sec:res}Results}
\subsection{Decision Diagram Representation of Quantum States}
Our quantum state preparation algorithm works efficiently by making use of a data structure named decision diagram (DD). Here we give a brief introduction to DDs and how they can be used to represent quantum states.

\begin{figure}
    \centering
    \subfloat[]{
    \includegraphics{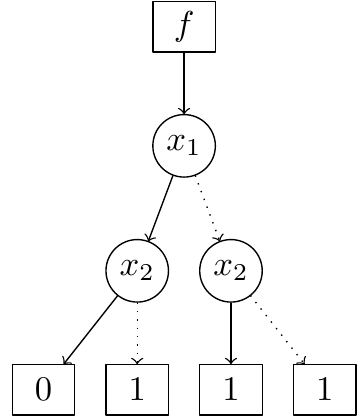}
    }
    \qquad
    \subfloat[]{%
    \includegraphics{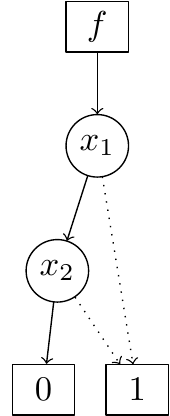}

    }
    \caption{The decision tree and the decision diagram for $f=x_1\overline{x_2} + \overline{x_1}x_2 + \overline{x_1}\overline{x_2}$. (a) Binary decision tree. (b) Binary decision diagram.}
    \label{fig:dtree_dd}
\end{figure}

\medskip
\noindent\textbf{Binary decision tree.}~ A binary decision tree is a rooted, directed, acyclic graph that represents a Boolean function $f=f(x_1,x_2,\dots, x_n)$. It consists of a \emph{root} node, several \emph{internal} nodes and several \emph{terminal} nodes. The root, usually printed as a square labelled $f$, features the start of the tree. The terminal nodes are labeled 0 and 1. The internal nodes, labeled $x_1$, $x_2$, $\dots$, $x_n$, represent the variables of $f$. Two adjacent internal nodes $x_1$ and $x_2$ are connected by a solid (dotted) arrow called edge to represent that the parent node $x_1$ (i.e., the node above) evaluates to 1 (0), and the node $x_2$ is called the one-child (zero-child) of $x_1$.  A terminal node $t$ has no children and is labeled 1 or 0 depending on the value of $f$ when its variables are evaluated to the values on the path that contains $t$.
Fig.~\ref{fig:dtree_dd}.a shows a binary decision tree for the Boolean function $f=x_1\overline{x_2} + \overline{x_1}x_2 + \overline{x_1}\overline{x_2}$.

%Given values of $a$ and $b$ shows a path from the function node to a terminal node. Along all paths, the variables are in the same order. In this figure $a$ and $b$ represent \emph{most significant bit} and \emph{least significant bit}, respectively.
\medskip

\noindent\textbf{Binary decision diagram.}~ A binary decision diagram can be obtained from a binary decision tree by applying a \emph{reduction} process, following the rules below:
\begin{enumerate}
    \item Two nodes are merged and their incoming edges are redirected to the merged node, if $i)$ they are both terminal and have the same value, or $ii)$ they are both internal and have the same sub-graphs.  
    \item An internal node is eliminated, if its two edges point to the same child. After elimination, its incoming edges are
    redirected to the child. 
\end{enumerate}

It is worth mentioning that the reduced tree is also called \emph{Reduced Ordered Binary Decision Diagram} (ROBDD) but is commonly referred to as BDD for simplicity. Fig.~\ref{fig:dtree_dd}.b shows the BDD obtained from the decision tree in Fig.~\ref{fig:dtree_dd}.a. First, three terminal nodes with value 1 merge to one. Next, node $b$ on the right-side of the tree eliminates as both children are terminal node 1.  

\medskip

\noindent\textbf{Algebraic decision diagram.}~ An \emph{Algebraic Decision Diagram} (ADD) is the same as a BDD, except that its terminal nodes can have any values  \cite{bahar1997algebric}. In other words, BDDs are ADDs whose terminal nodes have binary values. We can still apply reduction rules and get a ROADD, called ADD for short.

\medskip

\noindent\textbf{Quantum states represented by DDs.}~ Rather straightforwardly, an arbitrary $n$-qubit quantum state $|\varphi \rangle=\sum_{s\in S}\alpha_s|s\rangle$ can be represented by a decision diagram: for any $s\in S$, represent $s$ by a path in the tree and set its internal nodes to the qubit registers $q_1,q_2,\dots, q_n$, its edges to solid or dashed lines depending on the state of the registers, and its terminal node to $\alpha_s$. We then simplify the decision tree by removing all the paths corresponding to $s\not\in S$ and terminal nodes whose values are zero. Next, we further apply the reduction rules to get a ROADD (called ADD for short).
When the state is uniform, i.e., all the amplitudes are equal, the ADD can be simplified to a BDD, where a terminal node with the binary value 1 indicates that the associated paths have non-zero amplitudes. Each path $p$ of the reduced DD corresponds to one or more basis states $s\in S_p$, which is a subset of $S$. Denoting by $P$ the set of the paths of the reduced DD, the state to prepare can be recast in the form:
\begin{align}\label{reduced-state}
|\varphi\rangle=\sum_{p\in P}\sum_{s\in S_p}\alpha_{s}|s\rangle.
\end{align} 

Notice that all basis states $s\in S_p$ have the same amplitude.

\begin{figure*}

    \centering
    \subfloat[]{
    \includegraphics{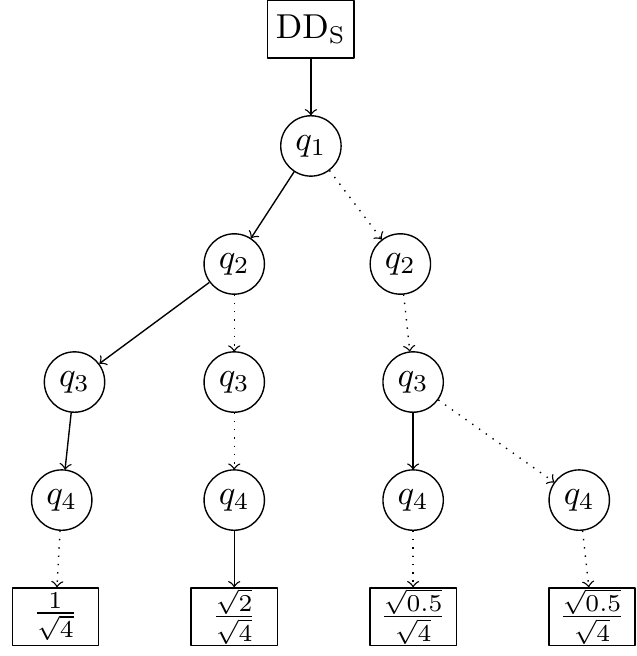}
    }
    \qquad
    \subfloat[]{%
   \includegraphics{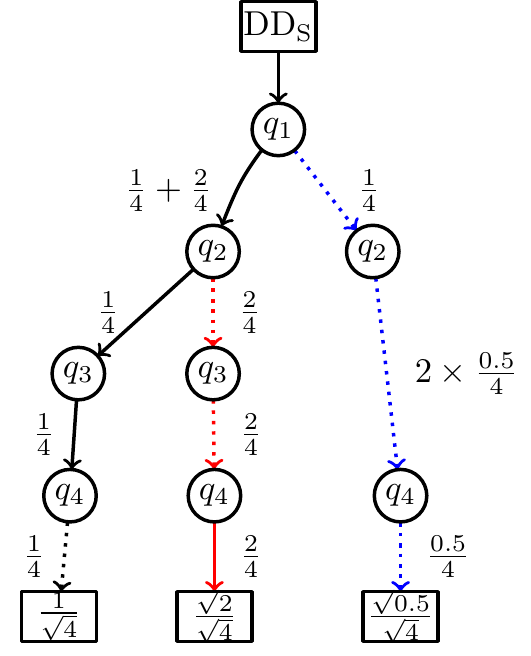}
}
    \caption{Decision diagram representation of the quantum state in the Example~\ref{exp:qs}. (a) Before applying reduction rules. (b) After applying reduction rules.}
    \label{fig:exp_dd}
\end{figure*}

\begin{example}
\label{exp:qs}
The 4-qubit state
\begin{equation}
    |\varphi \rangle = \frac{1}{\sqrt{4}} (|1110\rangle + \sqrt{2}|1001\rangle + \sqrt{0.5} |0010\rangle + \sqrt{0.5} |0000\rangle)
\end{equation}
 has index set $S = \{1110, 1001, 0010, 0000\}$ and non-zero amplitudes $ \{\frac{1}{\sqrt{4}}, \frac{\sqrt{2}}{\sqrt{4}}, \frac{\sqrt{0.5}}{\sqrt{4}}, \frac{\sqrt{0.5}}{\sqrt{4}}\}$. 
It can be represented by the decision diagram in Fig.~\ref{fig:exp_dd}.a. We represent each $s\in S$ with a binary string of qubits $q_1 q_2 q_3 q_4$ where $q_1$, $q_2$, $q_3$, and $q_4$ are internal nodes. Each path shows a basis state $s$, and the terminal node connecting to each path shows its corresponding amplitude. For example, $\{s_1=1110, \alpha=\frac{1}{\sqrt{4}}\}$ expresses that we have a path in which $\{q_1=1, q_2=1, q_3=1$,$q_4=0\}$ that connects to the terminal node $\frac{1}{\sqrt{4}}$.
Further noticing that on the right-side of the diagram (Fig.~\ref{fig:exp_dd}.a), two terminal nodes are equal which results in merging them. Furthermore, both left and right sub-graphs of $q_3$ are equal, so this node can be eliminated. Therefore, the decision diagram can be reduced to the ADD in Fig.~\ref{fig:exp_dd}.b which contains 3 paths instead of 4. Actually, the last two basis states $s_3=0010$ and $s_4=0000$ correspond to the same path $\{q_1=0, q_2=0, q_4=0\}$.

\end{example}

\subsection{DD-based Algorithm for Quantum State Preparation\label{ddAlgorithm}}

In this section, we present our DD-based algorithm for quantum state preparation. 
We assume that the quantum state to prepare is represented by either an ADD or a BDD (when it is uniform). Using DDs helps us to already have a quantum state without redundancies, which reduces the circuit cost. 

Our algorithm works by preparing the paths in a DD one-by-one.
For any $n$-qubit quantum state to prepare, our algorithm uses only one additional qubit $q_A$ as an ancilla, whose value is tagged $|{\rm yes}\rangle$ (regarded as $|0\rangle$ when used as a control qubit) or $|{\rm no}\rangle$ (regarded as $|1\rangle$ when used as a control qubit). Intuitively, $q_A$ serves as an indicator for whether a path has been created in the course of the state preparation. Paths that have been created are marked by $q_A\mapsto|{\rm yes}\rangle$ and, by using $q_A$ as control, we can avoid disturbing the created paths when creating a new path.
 
Each target-qubit in our quantum state preparation, transform $|0\rangle$ to a superposition of $\alpha|0\rangle+\beta|1\rangle$, where $|\alpha|^2$ ($|\beta|^2$) shows the probability of being zero (one) after measurement. To achieve this transformation, for some nodes, we need to apply a gate, called $G$, which is explained later. Therefore, we traverse the DD twice: 1) to compute the $G$ gate for each node, and 2) to prepare the quantum state.
\medskip

\noindent\textbf{Post-order traversal to compute $G$ gates.\label{gGate}}~ We traverse DD in post-order traversal (i.e. visiting one-child, zero-child, and parent nodes). For each node, we compute the probability of being one or zero from its corresponding one-child and zero-child. To compute zero probability (called $p_0$), for each node, we compute its portion from one-child (called $t_1$) and zero-child (called $t_0$) and then it equals to
\begin{equation}
    p_0 = \frac{t_0}{t_1+t_0}.
\end{equation}

As an example, consider the state in Fig.~\ref{fig:exp_dd}.b, post-order traversal results in first visiting $q_4$ in the left-side. The portion from one-child is 0 and from zero-child is $|\frac{1}{\sqrt{4}}|^2$ (as it is amplitude we need to square it). Hence, the probability of being zero equals to $\frac{\sfrac{1}{4}}{0+\sfrac{1}{4}}=1$. Next, we go through the upper node $q_3$, the portion from one-child comes from the summation of one-child and zero-child portions of $q_4$ which is $0+\frac{1}{4}$. The portion from zero-child is 0 and the zero probability is $\frac{0}{\sfrac{1}{4}+0}=0$. By continuing this procedure we obtain $t_1$ and $t_0$ which are written in the figure on the edges. 
Note that we need to consider the effect of eliminated nodes. If $e$ nodes are eliminated along an edge, the portion is multiplied by $2^e$. For example, in the right-side of the Fig.~\ref{fig:exp_dd}.b, on the zero-child of $q_2$, one node ($q_3$) is removed which results in $t_0=2^1\times \frac{0.5}{4}$.
Finally, $\alpha$ and $\beta$ for $G$ gates are computed by $\sqrt{p_0}$ and $\sqrt{1-p_0}$, respectively which we show it as 
\begin{equation}\label{G-p}
    G(p_0)|0\rangle = \sqrt{p_0}|0\rangle+\sqrt{1-p_0}|1\rangle.
\end{equation}

The above $G(p_0)$ can be implemented as a Pauli-$y$ rotation: $G(p_0)=R_y(2cos^{-1}(\sqrt{p_0}))$. 
 
\medskip

\noindent\textbf{Pre-order traversal to prepare the quantum state.}~ The algorithm begins with an empty quantum circuit and all qubits initiated as:
\begin{equation}
    |no\rangle_{q_A} \:\otimes \: |0\rangle_{q_1} \: |0\rangle_{q_2} \: ... |0\rangle_{q_n}.
\end{equation}

Starting from the root, the algorithm traverses the DD with pre-order traversal (i.e. visiting
parent, one-child, and zero-child nodes).
To accomplish the traversal, we need to define a pointer \emph{current\_node} that points to the current node we are working on. To navigate through the DD, we define functions \emph{one\_child} and \emph{zero\_child} which return child of the current node regarding solid and dotted edges, respectively.
While traversing through the DD, we compile the state preparation circuit according to the following rules:
\begin{enumerate}
    \item \textbf{Preparation.} If the current node $q$ is an internal node that is already on a path $p_i$, we do as follows. 
    \begin{itemize}
        \item If $q$ is a \emph{branching node}, which means it has both a zero-child and a one-child, we apply to the quantum circuit, a 2-controlled $G(p_0)$ gate [cf.~ Eq.~(\ref{G-p})] on $q$ with $q_A$ and the last node on the path that has a one-child as control qubits, where the value of $p_0$ is determined by the post-order traversal.
        Otherwise, $q$ either has a one-child or a zero-child. For the former case, we add a 2-controlled NOT gate on $q$ with $q_A$ and the last node on the path that has a one-child as control qubits. For the later case, we do nothing. 
        
        \item In addition, we need to consider the effect of reduced nodes between node $q$ and its children. A node is reduced when both its one-child and zero-child point to the same thing. Hence, the qubit with half probability is zero and with half probability is one. If this is the case, we append to the quantum circuit,  2-controlled G($\frac{1}{2}$) gates on reduced nodes with $q_A$ and the last node on the path that has a one-child as control qubits.
        
        \item If $q$ is the parent of the $i$-th terminal node, then we add a 2-controlled phase gate on $q$ with $q_A$ and the last node on the path that has a one-child as control qubits that adds a phase $e^{i\arg(\alpha_i)}$ to the path state $|s_i\rangle$.  
        
    \end{itemize}
    \item \textbf{Computing the ancilla.} If the current node is a terminal node, it means that we have prepared the current path. Hence, we need to compute the ancilla qubit to mark that the current path is prepared. We append to the quantum circuit a multiple-controlled NOT gate on $q_A$ with all qubits at branching nodes on path $p_i$ being control.
\end{enumerate}
\begin{algorithm}[h!]
\footnotesize
\caption{Deterministic Preparation of Quantum States using DD.}
\label{alg:general_idea}
\KwIn{DD representation of an $n$-qubit quantum state $|\varphi\rangle = \sum_{i=1}^{m} {\alpha_{s_i}} |s_i\rangle$, and $p_0$ values corresponding to each node of DD.}

%The classical specification of the set $S$ consisting of basis states $s_i$ and their amplitudes $\alpha_{s_i}$}
\KwOut{The quantum circuit $qc$ that prepares the desired quantum state.}
 
    Create a quantum circuit $qc$ with $n+1$ qubits corresponding to $q_A q_1 q_2...q_n$.
    
    Initialize the $qc$ with $|1\rangle_{q_A} \:\otimes \: |0\rangle_{q_1} \: |0\rangle_{q_2} \: ... |0\rangle_{q_n}$.\\
    
    Initiate the pointer $current\_node$ as the root of DD. %$v \leftarrow r$
    %\SetKwProg{Pn}{Proc}{:}{\KwRet}\\
    
    \textbf{PreOrder\_traversal $(current\_node, qc, p_0\_values):$}

   \If{$current\_node$ is a terminal node}{
    Append to $qc$ a multiple-controlled NOT gate with the qubits of $branches$ being controls and $q_A$ being the target.\\
        \textbf{return}
    }
   
    \If{$current\_node$ is a branching node}{
        Append to $qc$ a 2-controlled gate $G(p_0)$ [cf.~Eq.~(\ref{G-p}), with $p_0$ from $p_0\_values$ corresponding to $current\_node$] gate targeting the qubit corresponding to $current\_node$ controlled on ancilla qubit and the qubit corresponding to the last $|1\rangle$ in the path.\\
        
        Append the qubit corresponding to $current\_node$ and its value to $branches$.
        }
    \Else{
        \If{$ one\_child(current\_node) \neq nullptr$}{
            Append to $qc$ a 2-controlled NOT gate targeting the qubit corresponding to $current\_node$ controlled on ancilla qubit and the qubit corresponding to the last $|1\rangle$ in the path.
            
        }
            
    }
    \If{Some qubits are reduced between $current\_node$ and $one\_child(current\_node)$}{
        Append to $qc$ 2-controlled G($\frac{1}{2}$) gates targeting the reduced qubits with ancilla qubit and the qubit corresponding to the last $|1\rangle$ in the path as controls.
    }
    \If{$one\_child(current\_node)$ is a terminal node}{
     Append to $qc$ a 2-controlled phase gate that adds a phase $e^{i\alpha}$ (corresponding to this path) targeting the qubit corresponding to $current\_node$ controlled on ancilla qubit and the qubit corresponding to the last $|1\rangle$ in the path.
    }
    
    \textbf{PreOrder\_traversal $(one\_child(current\_node), qc, p_0\_values)$}
    
    \If{Some qubits are reduced between $current\_node$     and $zero\_child(current\_node)$}{
        Append to $qc$ 2-controlled G($\frac{1}{2}$) gates targeting the reduced qubits with ancilla qubit and the qubit corresponding to the 
        last $|1\rangle$ in the path as controls.
    } 
    
     \If{$zero\_child(current\_node)$ is a terminal node}{
     Append to $qc$ a 2-controlled phase gate that adds a phase $e^{i\alpha}$ (corresponding to this path) targeting the qubit corresponding to $current\_node$ controlled on ancilla qubit and the qubit corresponding to the last $|1\rangle$ in the path.
    }
    
      \textbf{PreOrder\_traversal  $(zero\_child(current\_node), qc, p_0\_values)$}

\end{algorithm}

 This is a recursive traversal where we visit the current node, one-child and zero-child, respectively. In other words, we prepare paths from the largest ($p_1$) to the smallest ($p_k$). In this way, we can order the elements of $P$ as $p_1\succ p_2\succ\cdots\succ p_k$. As a result, using this traversal we can prepare basis states in $S$ from the largest ($s_1$) to the smallest ($s_m$). The pseudo-code of the proposed algorithm is shown in Algorithm~\ref{alg:general_idea}. Note that, in the post-order traversal, we have already computed $p_0$ values of $G$ gates corresponding to each node and here we pass it as an argument to the algorithm. Line 5 of Algorithm~\ref{alg:general_idea} shows the applying rule 2: Computing the ancilla, and lines 8, 11, 14, 16, 19, and 21 illustrate different conditions of rule1: Preparation. Additionally, we recursively visit one-child and zero-child in lines 18 and 23.
 
\begin{figure*}
    \centering
    \scalebox{0.95}{\includegraphics{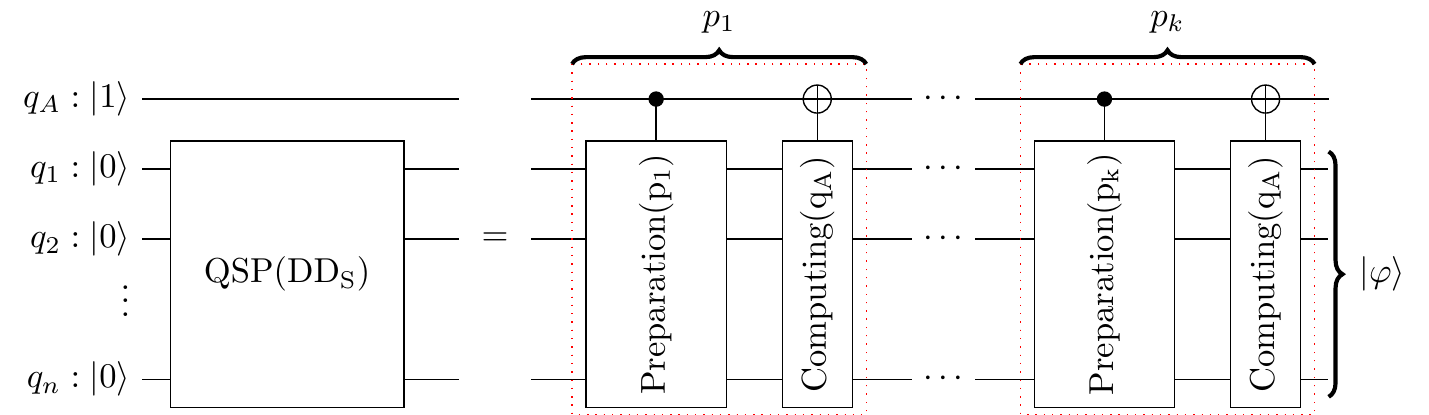}}
 \caption{The general structure of the quantum circuit for QSP over $\mathrm{DD_S}$.}
 \label{fig:general_idea}
\end{figure*}

Fig.~\ref{fig:general_idea} shows the general structure of the output quantum circuit of our algorithm. Note that for preparing $p_1$, the ancilla qubit is not needed, because there is no other path prepared before $p_1$. Moreover, as $p_k$ is the last path to prepare, we do not need to compute the ancilla qubit.

\begin{figure}
    \centering
 \includegraphics{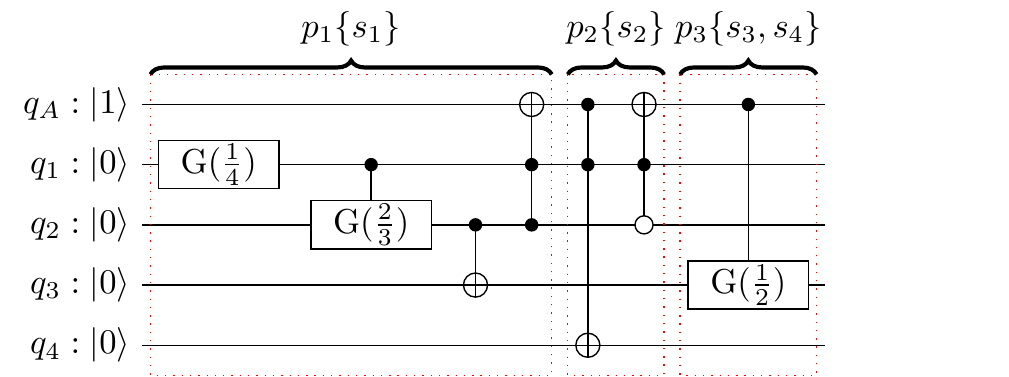}
 \caption{The generated quantum circuit for preparing the state presented as DD in Fig.~\ref{fig:exp_dd}.b.}
 \label{fig:exp_qc}
\end{figure}

\begin{example}
In this example, we show how to create a quantum circuit to prepare the state represented in Fig.~\ref{fig:exp_dd}.b. Pre-order traversal helps us to go through three paths presented by black, red, and blue colors. To compute $p_0$, values of $t_0$ and $t_1$ are shown in the figure.
Starting from the root, we need to append a  $G(\frac{1}{4})$ gate on $q_1$ that shows the probability of being zero for this qubit. Going through the black path $(p_1)$, on the next node $q_2$ there exists a branch which requires a 1-controlled $G(\frac{2}{3})$ gate. This is the first basis state and we do not need to check the ancilla qubit. Next, on $q_3$ there is not any branch but it has a one-child, so it is required to append a CNOT gate with the last $|1\rangle$ in the path $(q_2)$ as control. Next, for $q_4$ there is not any branch and there is only a zero-child that does not require any action. To compute the ancilla qubit, we need to add a multiple-controlled NOT gate on the ancilla qubit with 2 controls on branching nodes which are $q_1=1$ and $q_2=1$.

Afterward, the traversal returns to $q_2$ and goes through the red path $(p_2)$. It goes to $q_3$, there is not any branch and there is only a zero-child that does not require any action. Next, $q_4$ has a one-child and so we need to add a 2-controlled NOT gate on $q_4$ with ancilla and $q_1$ which is the last $|1\rangle$ in the path as control qubits. Then, to mark that $p_2$ is prepared, we add a 2-controlled NOT gate on the ancilla qubit with $q_1=1$ and $q_2=0$.

Finally, the algorithm goes back to the root again and traverses the blue path $(p_3)$. $q_2$ has a zero-child and we do not need to add any gate for it. Next, the $q_3$ is removed which requires adding a G$(\frac{1}{2})$ gate that shows with the half probability it is zero. There is not any last $|1\rangle$ in this path so it only has one control which is the ancilla. Then, $q_4$ has zero-child and again we do not need to add any gate for it. Note that reduced node $q_3$ here help us to prepare $s_3$ and $s_4$ together. This reduces the number of iterations and so circuit cost. Moreover, as this path corresponds to the last basis states $s_3, s_4$, we do not need to compute the ancilla qubit. Fig.~\ref{fig:exp_qc} shows the generated quantum circuit.
\end{example}

 
\subsection{Numerical Experiments}
In this section, we evaluate the proposed algorithm over the state of the art~\cite{sparseQSP}. Our algorithm is implemented in an open-source tool, called \emph{angel}\footnotemark[1]. All experiments are conducted on an Intel Core i7, 2.7 GHz with 16 GB
memory. 
\footnotetext[1]{A C++ library for quantum state preparation, \href{https://github.com/fmozafari/angel}{https://github.com/fmozafari/angel}.}
\medskip

\begin{figure*}
\centering
\subfloat[]{ 
    \includegraphics{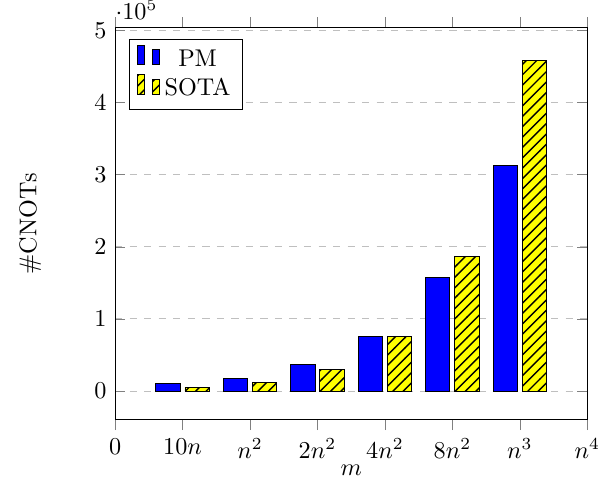}
     }
     \subfloat[]{ 
     \includegraphics{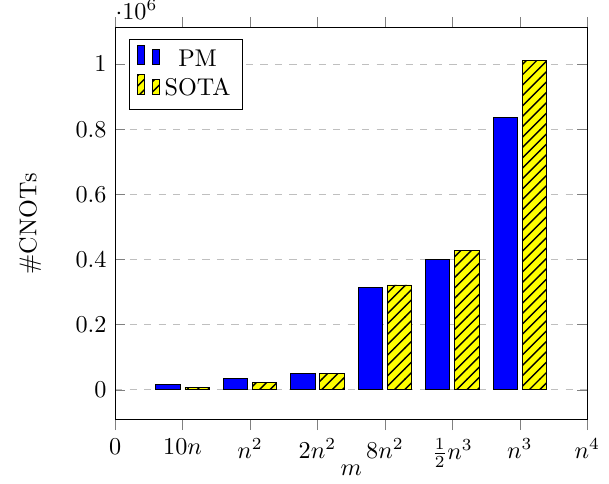}
     }
     
     \subfloat[]{ 
    \includegraphics{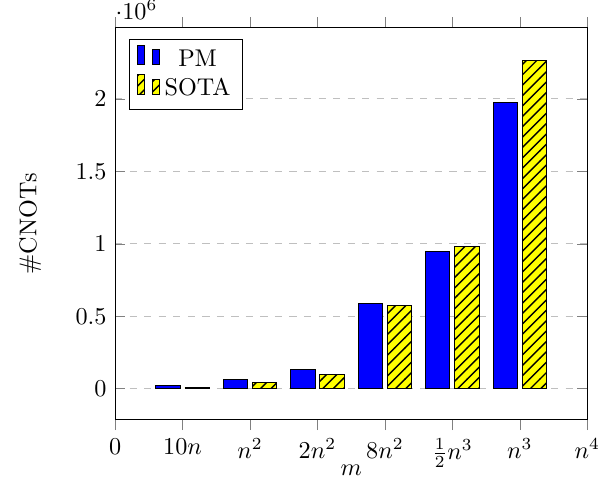}
     }
     \subfloat[]{ 
    \includegraphics{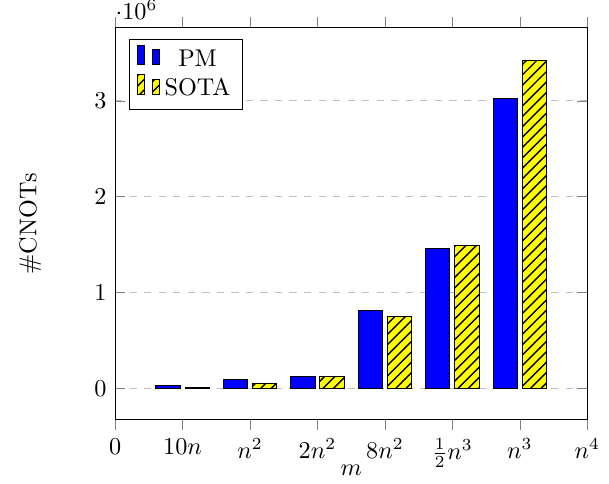}
     }
     
\caption{{\bf Comparison between the CNOT complexities of our proposed method (PM) and the state-of-the-art (SOTA) method.} Our PM is compared to the best-known algorithm (STOA) in \cite{sparseQSP} on random sparse states of $n$ qubits. For different $n$, we plot the number of CNOT gates required in both algorithms as a function of $m$, the number of non-zero amplitudes. It can be seen that PM requires fewer CNOTs in the interval between $2n^2$ and $n^3$ for $n=16, 20$, and $8n^2$ and $n^3$ for $n=25, 28$. Moreover, the more increasing of $m$ results in the more reduction of CNOTs. (a) $n = 16$. (b) $n = 20$. (c) $n = 25$. (d) $n = 28$.}
        \label{fig:cnots_difk}
\end{figure*}

\noindent\textbf{Random states.}~ We evaluate our algorithm on randomly generated states with different amplitudes. The parameter $m$ denotes the number of basis states with non-zero amplitudes. We change $m$ depending on $n$ with different degrees. We compare the size of the circuits produced
by our proposed method (PM) with the state-of-the-art method (SOTA) presented in~\cite{sparseQSP}. The final circuits consist of CNOTs and single-qubit gates as elementary quantum gates. We only consider the number of CNOTs as they are more expensive than single-qubit gates in the NISQ. But consider that reducing CNOTs means we are reducing single-qubit gates as well. Fig.~\ref{fig:cnots_difk} shows results for $n=16$, $20$, $24$, and $28$. For each combination of parameters shown in the figure, we sampled 10 random
states and show the average values.
Each sub-figure shows how the number of CNOTs grows as we increase $m$ as a function of $n$. For small $m$, SOTA is better as it is an efficient idea for sparse states. But by increasing $m$ our results closes to SOTA and finally for $m=n^3$, PM outperform SOTA up to 31.85\%, 17.4\%, 13.1\%, and 11.4\% for $n$ equal to 16, 20, 25, and 28, respectively. The reason is that in the decision diagram representation, for large $m$, there is a better sharing between basis states which results in a sparse decision diagram. The results for $n=16$ are better than those for larger values of $n$ because the percentage of non-zero amplitudes is higher for $n=16$.
Considering the sparsity condition in~\cite{sparseQSP}, $m \in o(\frac{2^n}{n})$, these values of $m$ are still sparse. We conclude that our method is more useful than SOTA for large $m$.

\medskip

\noindent\textbf{\label{prop:exp}Special states.}~ To show our improvement for small $m$, we extract special states whose DD representations are sparse and the reduction rules work well on them. These states are mostly uniform states that share paths better. These states benefit from the effect of reduced nodes which reduce the number of paths and branching nodes in each path. This results in reducing the number of multiple-controlled gates and their control qubits which is required for computing the ancilla qubit. Table~\ref{tab:res_specific} shows the average results for such states (set 1, 2, 3, and 4) in comparison with SOTA. We consider two different numbers of qubits 20, 30, and small $m=n, 10n$. For each quantum state set, using the proposed method, we extract results regarding the number of nodes, number of reduced nodes, number of paths, and number of CNOTs. The number of reduced nodes shows that we can prepare several basis states together which reduces the number of CNOTs. Moreover, the number of paths, which is important in our complexity, is much less than the number of basis states, which results in reducing CNOTs. We also extracted the number of CNOTs by SOTA. Comparison shows that we reduce the number of CNOTs up to 98\%.

Quantum Byzantine agreement (QBA) represents the quantum version of Byzantine agreement which works in constant time. In this protocol, for $n$ players, we need to prepare the quantum state
\begin{equation}
    |\varphi\rangle = \frac{1}{\sqrt{n^3}} \sum_{i=1}^{n^3} |i\rangle
\end{equation}
on $n$ qubits. For large $n$, this state is sparse. Table~\ref{tab:res_specific} shows its results. The proposed method prepares this state more efficiently. As shown in the Table~\ref{tab:res_specific}, we reduce number of CNOTs by 99.97 \% for QBA when $n=30$. The reason is that the number of paths is much less than the number of non-zero basis states.

\begin{table*}[t]
\centering
	\caption{Experimental results for quantum states (qs) that have a sparse DD.}
	\renewcommand{\arraystretch}{0.9}
	\label{tab:res_specific}
    \setlength{\tabcolsep}{2pt}
  \begin{tabularx}{1\textwidth}{XlXlXrrrrXrXr}
    \toprule
          &    &  &   &  & \multicolumn{4}{c}{PM}                       &  & \multicolumn{1}{c}{SOTA} &  &           \\
         \cmidrule(lr){6-9}
    	\cmidrule(lr){11-11}
    qs & $n$  && $m$   &  & \#nodes & \#ReducedNodes & \#paths ($k$) & \#CNOTs &  & \#CNOTs                  &  & Imp. (\%) \\
    \midrule 
set 1      & 20  &  & $n$   &  & 33      & 6              & 2       & 13      &  & 275                      &  & 95.27     \\
set 2      & 20  &  & $10n$ &  & 41      & 25             & 5       & 190     &  & 9983                     &  & 98.10     \\
set 3      & 30  &  & $n$   &  & 60      & 10             & 4       & 62      &  & 463                      &  & 86.61     \\
set 4      & 30  &  & $10n$ &  & 78      & 41             & 9       & 568     &  & 17019                    &  & 96.66     \\
QBA    & 20  &  & $n^3$ &  & 32      & 110            & 18      & 1165    &  & 1361456                  &  & 99.91     \\
QBA    & 25  &  & $n^3$ &  & 37      & 123            & 19      & 1321    &  & 2974248                  &  & 99.95     \\
QBA    & 30  &  & $n^3$ &  & 44      & 141            & 22      & 1591    &  & 5512726                  &  & 99.97    \\
    \bottomrule	
  \end{tabularx}
\end{table*}

\subsection{Algorithm Performance}
\noindent\textbf{Correctness.}~ First we explain how our algorithm prepares an arbitrary $n$-qubit state, given by Eq.~(\ref{reduced-state}), without any approximation error. It is enough to show that, starting from the initial state $|{\rm no}\rangle_{q_A}\otimes|0\rangle^{\otimes n}$, in each iteration, in which the path $p_i\in P$ is traversed, we create a part 
$|{\rm yes}\rangle\otimes \sum_{s\in S_{p_i}}\alpha_{s}|s\rangle$ of the target state, where $S_{p_i}$ is the collection of basis states that are merged into path $p_i$ in the creation of DD.

Meanwhile, we keep the prepared parts $|{\rm yes}\rangle\otimes\sum_{j<i}\sum_{s\in S_{p_j}}\alpha_{s}|s\rangle$ untouched. (Be reminded that $p_1\succ p_2\succ\cdots\succ p_k$.) In this way, after traversing the last path $p_k$, we end up with $|{\rm yes}\rangle\otimes\sum_{j=1}^k\sum_{s\in S_{p_j}}\alpha_{s}|s\rangle$ as desired, where the system is in the target state and is uncorrelated with the ancillary qubit $q_A$.

To see how this is achieved in each iteration, first, notice that a path is uniquely characterised by its branching nodes and their values. For example, the path $000101$ can be specified by $q_1=0$, $q_4=1$, $q_5=0$, and $q_6=1$, as in between $q_1$ and $q_4$ we adopt the convention that both $q_2$ and $q_3$ take the same value as $q_1$. Therefore, it is enough to prepare a branch without altering other branches, by acting on each node using its preceding branching nodes as the control. In our algorithm (more precisely, in preparation rule), we further reduce the cost by the following crucial observation: When working on a qubit $q$ in $p_i$, consider its closest ancestor whose value is one in $p_i$, denoted by $\tilde{q}$.
Since the sequence $p_1,p_2,\dots,p_{k}$ is also ordered, only those completed parts (i.e.~the partial state $\sum_{j<i}\sum_{s\in S_{p_j}}\alpha_{s}|s\rangle$) corresponding to paths $p_1,\dots,p_{i-1}$ can have $\tilde{q}=|1\rangle$. On the other hand, for those paths where $\tilde{q}=|1\rangle$, they have already been completed and thus are tagged $|{\rm yes}\rangle$ (regarded as $|0\rangle$ when used as a control qubit) on $q_A$. Therefore, it is sufficient to use two qubits ($\tilde{q}$ and $q_A$) as the control to make sure that other completed parts are unaltered in the course of preparing the $i$-th part. As a result, we can complete the $i$-th part without affecting the prepared paths by following preparation rule of the algorithm. Since the branching nodes uniquely determine a path, we can flip the value of $q_A$ of the $i$-th part from $|{\rm no}\rangle$ to $|{\rm yes}\rangle$ by following the computing the ancilla rule.

\medskip

\noindent\textbf{\label{prop:opt}Circuit complexity.}~
In DD, $p_i$ and $p_{i-1}$ may share a common sub-path; therefore, we do not need to start preparation from the root for every $p_i$. This helps us to append fewer gates and reduces the number of CNOTs and single-qubit gates.

Our idea is based on DD which we use reduced ordered BDD or ADD to represent the quantum state. Using them allows us to have a compact representation for the state and to remove redundancies that reduces circuit cost in the preparation. Moreover, reduced nodes help us to prepare some basis states together. Hence, in contrast to the previous works that the number of basis states ($m$) is considered in the circuit complexity, the number of paths ($k$) is important in our complexity, and always
\begin{equation}
    k \leq m.
\end{equation}
 
According to the subsection~\ref{ddAlgorithm}, preparing a path is divided into two parts: preparing the path and computing the ancilla qubit. As a quantum circuit, it requires a sequence of 2-controlled gates to prepare the corresponding basis state (or basis states), and a multiple-controlled NOT gate to compute the ancilla qubit. 

To compute the circuit complexity, we need to compute the number of 2-controlled gates for the first part, and the number of controls for the second part. The number of 2-controlled gates depends on the number of branching nodes in the path, and the number of one-child in the path of the corresponding basis state. %As we have an sparse state the number of branching nodes could not be close to $n$ for all basis states.
Moreover, in the DD, paths have overlap and we prepare each basis state from the last common node with the previous basis state instead of starting from root. Considering this optimization, our algorithm reduces the number of 2-controlled gates. But in the worst-case we require $n$ 2-controlled gates. Decomposition of each 2-controlled gates require 4 or 6 CNOT, and so we need $O(n)$ CNOTs. For the second part, the number of controls is equal to the number of branching nodes in the path. Then, we make use of the method proposed in~\cite{gidney} to decompose multiple-controlled NOT gate using $O(n)$ CNOT gates and one ancilla. We repeat same procedure for $k$ paths and so, in total, the number of CNOTs is equal to
\begin{equation}
    \#CNOTs = k \times O(n).
\end{equation}

%In addition, our method works better for the states that the number of ones in the basis state string is less. This is due to the fact that when there is not a branching node, if the qubit is zero we do not need to add a 2-controlled gate.

\medskip

\noindent\textbf{Time complexity.} We traverse DD twice to first compute $G$ gates and secondly prepare the quantum state. As we visit each node once, each traversal is linear in the number of nodes, and such a number increases mildly (but not always) with problem size (i.e., qubits). The number of nodes depends on the number of paths and the number of qubits in each path. Hence, the number of nodes is always less than $kn$ as there exist sharing nodes at least for the root. As a result, the classical runtime is less than $2kn$, which is less than the time required by the state of the art~\cite{sparseQSP}.
\medskip
\section{\label{sec:con}Discussion}
In this paper, we have proposed an algorithm to prepare quantum states deterministically. Our idea is based on preparing basis states one-by-one instead of operating one-by-one on the qubits. The latter is the key idea in general quantum state preparation algorithms. We have utilized DDs to represent quantum states in an efficient way. This allows our algorithm to be dependent on the number of paths where related works~\cite{de2021double, malvetti2021quantum,sparseQSP} are dependent on the number of basis states. We prepare the paths from the largest to the smallest regarding their binary bit strings. To do so, we traverse the DD in the pre-order traversal. Through this traversal, we visit nodes on a path. For each node, depending on the existence of its two children (i.e. \emph{branching node}), we decide to append either 2-controlled single-target gates with different targets or just skip that. Upon preparing the path, an ancilla qubit is computed by adding a multiple-controlled NOT gate with the number of controls equal to the number of branching nodes in the path. Considering the decomposition method in~\cite{gidney}, preparing each path and computing the ancilla qubit require $O(n)$ CNOTs. As a result, the final circuit cost depends on the number of paths and equals $O(kn)$. Using DDs helps us to have a compact representation of the state vector by reducing redundancies. The main advantages of our DD-based approach are:
\begin{itemize}
     \item For preparing each path, we do not need to start from the root node. We go back to the last common node with the previous path.
    \item When there are redundant nodes, removing them causes merging basis states to the same path and we can prepare them together. This helps in two ways. First, it reduces the number of iterations ($k$). Second, it reduces the number of branching nodes in paths which decreases the number of control qubits for computing the ancilla qubit.
\end{itemize}

Experimental results show that our idea works well for sparse DDs in which the number of paths and branching nodes are reduced. A sparse DD will be achieved when either $m$ is small or $m$ is not small but basis states share paths and can be prepared together. Hence, our algorithm besides SOTA, work very well to prepare sparse states and states with sparse DDs. As future work, we can consider variable reordering in DD to get a more sparse DD. 

As a concluding remark, we note that analyses in this work are done assuming full connectivity between qubits, whereas a realistic \textit{Quantum Processing Unit} (QPU) is often subject to limited qubit connectivity. In the following, we compare our algorithm to SOTA over an example that takes into account the limited qubit connectivity.

\begin{figure}
    \centering
 \includegraphics{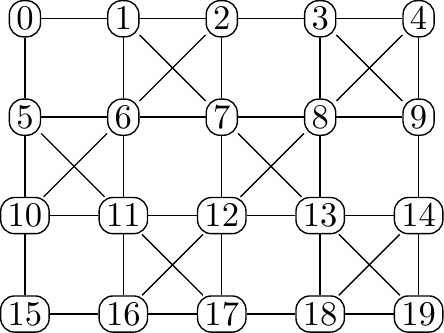}
 \caption{{\bf Coupling map for the IBM Q Tokyo.} Here $0,~1,~\dots,~19$ stand for physical qubits, and the edges indicate their connectivity.}
 \label{fig:ibmq20}
\end{figure}

\begin{example}
Consider preparing a uniform-amplitude quantum state corresponding to 
\begin{equation}
    S = \{1000, ~0100, ~0011, ~0010, ~0001, ~0000\}.
\end{equation} 
To prepare it on a QPU with full qubit connectivity, our method and SOTA require 10 and 12 CNOTs, respectively.
When preparing it on IBM's 20-qubit Tokyo  with a coupling map as shown in Fig.~\ref{fig:ibmq20}, the cost depends on the mapping from logical qubits to physical qubits, which we choose to be:
\begin{equation}
    \{q_1 \rightarrow 0,~ q_2 \rightarrow 1,~ q_3 \rightarrow 6, ~q_4 \rightarrow 7 \}, ~ q_A \rightarrow 2.
\end{equation}
Under this mapping, compiling the  circuit  generated by our method and compiling the one generated by SOTA both result in two extra SWAP gates. As each SWAP is decomposed into three CNOTs, the final numbers of CNOTs for our method and SOTA are 16 and 18, respectively. Hence, for this example, our method outperforms SOTA both before and after the compilation.
\end{example}

\medskip 
\noindent{\bf Code availability}~ 
The algorithm that we discussed in this paper is part of the \emph{angel} library (\url{https://github.com/fmozafari/angel}), in the path `include/angel/quantum\_state\_preparation/'. angel is a C++ open-source library for quantum state preparation.

\medskip

\noindent{\bf Acknowledgements}~
This research was supported by the Google PhD Fellowship, by the Natural Science Foundation of Guangdong Province (Project 2022A1515010340), and by HKU Seed Fund for Basic Research for New Staff via Project 202107185045. 

%\appendix

\bibliography{refs}

\end{document}